\documentclass[useAMS,usenatbib]{article}

\usepackage{times}
\usepackage{bm}
\usepackage{natbib}
\usepackage[figuresright]{rotating}
\usepackage{graphicx}
\usepackage{amssymb}
\usepackage{amsmath}

\newcommand{\Ee}{{{\rm E}}}

\newcommand{\bX}{\mbox{\boldmath $X$}}

\newcommand{\bR}{\mbox{\boldmath $R$}}

\newcommand{\dd}{{\rm d}}

\newcommand{\dinf}[1]{\longrightarrow_{#1}}

\newcommand{\wli}[1]{\mbox{$\longrightarrow \hspace{-0.60cm} {\bf /} \hspace{0.40cm}_{#1}~$}}

\newcommand{\N}{{{\cal N}}}
\newcommand{\X}{{{\cal X}}}

\newcommand{\LL}{{\cal L}}

\newcommand{\Ob}{{{\cal O}}}

\newcommand{\PP}{{{\cal P}}}

\newtheorem{Definition}{Definition}

\newcommand{\FF}{{{\cal F}}}

\usepackage{colortbl}

\title{Dealing with death when studying disease or physiological marker: the stochastic system approach to causality}

\author{Daniel Commenges
INSERM, U 1219, Bordeaux,  F33076, France}

\begin{document}





\label{firstpage}
\maketitle

\begin{abstract}
The stochastic system approach to causality is applied to situations where the risk of death is not negligible.
This approach grounds causality on physical laws, distinguishes system and observation and represents the system by multivariate stochastic processes. The particular role of death is highlighted, and it is shown that local influences must be defined on the random horizon of time of death. We particularly study the problem of estimating the effect of a factor $V$ on a process of interest $Y$, taking death into account. We unify the cases where $Y$ is a counting process (describing an event) and the case where $Y$ is quantitative; we examine the case of observations in continuous and discrete time and we give a typology of cases where the mechanism leading to incomplete data can be ignored. Finally, we give an example of a situation where we are interested in estimating the effect of a factor (blood pressure) on cognitive ability in elderly.
\end{abstract}

{\bf keywords:} ageing; causality; death; epidemiology; joint models; markers; stochastic system.

\section{Introduction}\label{sec:Introduction-DynCaus}
There are many epidemiological studies of risk factors of Alzheimer disease or dementia, or the decrease of cognitive function in the elderly. As has been identified by \cite{weuve2015guidelines}, one of the major methodological problem is that of selection of the sample due to death. It is not obvious to treat this problem, which may also arise in other studies of severe diseases, like cancer for instance.
We aim to investigate the issue of estimating the effect of a factor (which may be time-dependent) on a physiological state which can be binary (such as dementia) or quantitative (such as cognitive ability) in situations where the death risk is not negligible (such as in ageing studies). We have in mind situations where longitudinal observations of events and/or quantitative markers can be recorded. The question is not a pure statistical one, but the challenge is rather to formulate the problem correctly so as to identify the relevant parameters.
A similar problem was tackled by \citet{Rubin2006a} using potential outcomes and principal stratification; however the type of observations he considered are not longitudinal and do not allow identifying a dynamic model and do not even allow estimating reliably the estimand he proposed, the survival average causal effect (SACE). This estimand is itself questionable, but it is out of the scope of this paper to discus it. 

Our approach does not use potential outcomes but is based on the dynamical approach to causality. The dynamical approach to causality uses the formalism of stochastic processes and the concept of system. A general definition of causal influence was given by \citet{Aalen1987};  \citet{Didelez2008} developed this idea for marked point processes models and proposed influence graphs; a particular approach called ``dynamic path analysis'' has been developed by \citet{Fosen2006}. These authors do not have specifically studied the special role of death.
The core of the approach presented here is essentially based on \citet{Commenges2009}, and has been further developed in \citet{Gegout-Petit2010}, \citet{commenges2015stochastic} and in \citet{commenges2015dynamical} (Chapter 9), and we call it ``the stochastic system approach to causality''.
It allows  mixing counting processes and diffusion precesses, thus allowing us to develop insight in the so-called joint models where both events and markers are modelled. The concept of system is nearly absent in biostatistics but is ubiquitous in physics. A philosophical analysis of the concept of ``system'' can be found in \citet{Wimsatt1994}.

Our schedule is :
(i) to represent the problem in the stochastic system formalism;
(ii) to identify the relevant parameters to estimate when the risk of death is not negligible;
(iii) to look at the conditions on the observation scheme, including latent processes, for estimating these parameters;
(iv) to investigate in which cases some parameters can be estimated in a smaller system where ``death''  is not represented.

An originality of the presentation is that it unifies problems involving a binary state (usually treated by multistate models) and a quantitative state (usually treated as ``repeated measures''). We also present a general observation scheme which unifies observations of binary and quantitative processes including latent processes.
We recall the  background of the dynamic approach to causality in Section \ref{sec:background-dyncausal}. In Section \ref{ageing studies} we present some general considerations for taking death into account.
 In Section \ref{sec:observation} we study general schemes of observations and criteria for ignorability of these schemes. In Section \ref{effect-on-Y}, we tackle the problem of estimating the effect of a factor on a physiological state, in presence of death, examining when the mechanism leading to missing data is ignorable. In Section \ref{sec:incompletesystem} we examine what can be done with an incomplete system not including death. In Section \ref{sec:Illustration}, we develop the example of cognitive ability in the elderly. Section \ref{sec:conclusion} concludes.

\section{The dynamical approach to causality}\label{sec:background-dyncausal}

\subsection{Representation by stochastic processes}
The starting point of the dynamical approach is to consider that we have a better representation of phenomena and their causal relationship by using stochastic processes than by using ordinary random variables, and most of the time it is better to consider that these processes live in continuous time.
This leads us to a change of paradigm. In the  conventional paradigm, we observe random variables and we search for a model that will fit them; in the stochastic system paradigm, a system is represented by a stochastic process and we collect observations of the state of this system which will allow estimating the law of the process.
Thus, it is useful to distinguish between the model for the system and the model for the observation, a classical distinction in automatics  \citep{kalman1961new,jazwinski1970stochastic} but not in biostatistics.
To illustrate this different point of view and the further introduced concepts, we will take a toy example.
Consider first that we are interested in a physiological quantity, say blood pressure, and how it varies with age. We have observations $\tilde V_j$ of blood pressure at age $t_j$ and we may model them by $\tilde V_j=\beta_0+\beta_1 t_j+\varepsilon_j$ with some assumptions on the distribution of the $\varepsilon_j$'s: this is the conventional approach. In the stochastic system approach we have a model for the system which is a model for a stochastic process $V$ in continuous time; this gives justice to the fact that there is some blood pressure at any time. The law of the process can be given by its Doob-Meyer decomposition and a possible model is: $V_t=\beta_0 +\beta_1 t +\omega B_t$, where $B_t$ is a Brownian motion or
in differential form:
$$\dd V_t=\beta_1 \dd t +\omega \dd B_t,$$
which makes the dynamics of the process more visible. Observations $\tilde V_j$ are then noisy observations of the process $V$ at time $t_j$ : $\tilde V_j=V_{t_j}+\varepsilon_j$. One advantage of this formulation is that it gives a natural correlation structure for the $\tilde V_j$'s.

Suppose now that we are interested in the occurrence of a type of event, say dementia. Rather than modeling the distribution of the time of occurrence of the event, we can find the law of a counting process $Y$.
The law of $Y$ can be given by its Doob-Meyer decomposition $Y_t=\Lambda_t +M_t$, or its differential form $\dd Y_t=\lambda_t \dd t+\dd M_t$, where $\Lambda=(\Lambda_t)$ is the compensator and $\lambda=(\lambda_t)$ the intensity of the process. We may have continuous-time observations (with possibly right-censoring) or discrete-time observations (inducing interval-censoring).

For complex problems, we need multivariate processes. In the case there are several types of events, this could be represented by a multistate process; we will prefer a representation by a multivariate counting process because we are interested in the relation between the different components of the multivariate stochastic process. For instance we are interested in both dementia and death; the interaction of these two events can be represented by an illness-death model. Alternatively, this can be represented by a bivariate counting process (see Section \ref{pos-systems}).

The multivariate stochastic process $\bX$ can have components which are counting processes and others which are diffusion processes, allowing us to analyze the relationships between events and continuous phenomena, both typically evolving in continuous time. Suppose we are interested in both blood pressure and dementia, we can consider a joint model for the two processes:
$\dd V_t=\beta_1 \dd t +\omega \dd B_t$ ; $\dd Y_t=\lambda_t \dd t+\dd M_t$. The intensity of $Y$ can be modeled as: $\lambda_t=I_{{Y_{t-}=0}}\alpha_0(t)e^{\gamma V_t}$, where $\alpha_0(\cdot)$ is the baseline hazard function. We may have discrete-time observations of $V$ and continuous- or discrete-time observations of $Y$, allowing estimating the parameters of this model.

\subsection{Local influence in stochastic processes}
Given a system represented by a multivariate stochastic process $\bX$, a criterion of local independence  is defined in terms of measurability of processes involved in the Doob-Meyer representation \citep{Aalen1987}. \citep{Commenges2009} called the local independence WCLI (weak local conditional independence) because they also defined a criterion of strong local independence (SCLI); when WCLI does not hold, there is direct influence, when SCLI does not hold while WCLI holds, there is indirect influence. In short, if a component of the stochastic process $X_k$ does not appear in the compensator of the Doob-Meyer decomposition of $X_j$ we say that $X_j$ is WCLI of $X_k$. In our above example of a joint process $\bX=(Y,V)$ we have:
\begin{eqnarray*}
\dd V_t&=&\beta_1 \dd t +\omega \dd B_t \\
\dd Y_t&=&I_{{Y_t=0}}\alpha_0(t)e^{\gamma V_t} \dd t+\dd M_t,
\end{eqnarray*}
where the martingales $B$ and $M$ are orthogonal. 
We see that $V_t$ appears in the intensity of $Y$: if $\gamma \ne 0$, this intensity would not be {\em measurable} in a filtration\index{filtration} not including $V$. Equivalently  we could say that marginally to $V$, $Y$ does not have the same intensity. On the contrary $Y$ does not appear in the intensity of $V$: we do not need any information on $Y$ to know the dynamics of $V$. We shall say that $V$ is WCLI of $Y$, but that $Y$ is not WCLI of $V$.

Conversely, if a component of $\bX$, $X_k$, is not WCLI of another component, $X_j$, we say that  $X_j$ has a ``direct influence'' on $X_k$, and we note: $X_j \dinf{\bX} X_k$. In our example, we would note $V \dinf{\bX} Y$ and since $V$ is WCLI of $Y$ we can also note $Y \wli{\bX} V$. It is important to note that the direct influences depend on both the system $\bX$ and the probability law.

\subsection{Graphical representation}
A graph can then be constructed having the components of the stochastic process as nodes and directed edges where there are direct influences. This is analogous to classical graphical models\index{graphical models} with the difference that nodes are stochastic processes rather than random variables and the graph may be cyclic; in particular we may have both $X_j \dinf{\bX} X_k$ and $X_k \dinf{\bX} X_j$. An advantage of these process graphs is that they are  more concise than the conventional directed acyclic graphs (DAG) (based on random variables rather than stochastic processes), and also more concise than the graphs for multistate models.

\subsection{Perfect and NUC Systems, and causal influences}\label{causalinfluences}
How can the mathematical property of ``direct influence'' between components of a process under a particular probability $\PP$ be used for exploring causality? Answering this question may be possible if we have a definition of ``causal influence.'' We give a definition based on the concept of ``system'' universally used in physics and in automatics \citep{kalman1961new}; see philosophical aspects in \citet{Wimsatt1994}. It is postulated that, for a given ``level,'' there exists a sufficiently large system $\bX^M$ and  physical laws allowing to compute the true probability $\PP^*$ for events of interest. See a precise definition and examples on  gravitation law and mechanistic knowledge in HIV infection in \citet{Commenges2009} (Section 3.1).

 In our example on dementia we will try to represent all the processes that may influence dementia. Such a system denoted by $\bX^M$ will be called a ``perfect'' system for dementia. This allows defining ``causal influences''. A direct influence of a process $V$ on $Y$ (dementia) in $\bX^M$ under $\PP^*$ is called a direct ``causal influence''. The causal effects (which are the quantification of causal influences) can be summarized by different contrasts between the intensities obtained for different values of V.

Many issues in causality come from the fact that we generally do not work with $\bX^M$ but with smaller systems. If we do not have a perfect system for Y, by definition there is a process $U$ which influences $Y$ and which is not included in the system. Then the question is whether it is possible to estimate marginal causal effects of a factor $V$ on $Y$ in this system. Such process $U$ is a potential unmeasured confounder \citep{arjas2004causal}; it is a confounder for $(V,Y)$ if it influences $V$ in the larger system $(\bX, U)$. If it does not influence $V$, then $\bX$ is a system with no unmeasured confounders for $(V,Y)$, and marginal causal effects can be estimated. Such systems will be called ``NUC systems for (V,Y)''.
See Section \ref{pos-systems} for an illustration.

\subsection{Fixed and random horizon for WCLI}\label{randomhorizon}
In fact, WCLI and influences can be defined on a finite horizon, $\tau$. It could well be that a process has an influence on another process until a certain time only, or that we are not interested in the possible influence after a certain time. This horizon can be fixed or random.

A random horizon is particularly interesting when studying the effect of a process which represents a risk factor of a disease, because we are generally not interested in the effect of the disease on the risk factor.
As an example, consider a system with two processes $V$ and $Y$; $V$ could represent systolic blood pressure and $Y$ dementia; $V$ has a continuous state space, while $Y$ is a $0-1$ counting process. Epidemiologists are interested in knowing whether high blood pressure is a risk factor of dementia. In general they are not interested in the effect of dementia on blood pressure, although such an effect is not excluded. Another example would be the relation between tobacco consumption ($V$) and lung cancer ($Y$). Epidemiologists are interested in the effect of tobacco on cancer. There may well be an effect of cancer on tobacco consumption since people with cancer are likely to stop smoking, but this is not of primary interest for epidemiologists.

So we are interested in knowing whether $Y$ is WCLI of $V$, on $(0, T_Y)$, where $T_Y$ is the time of occurrence of the event (dementia or cancer). After $Y$ has jumped, the intensity is null, so that $V$ cannot influence $Y$ after the jump.  If we stop at $T_Y$, there can be no effect of $Y$ on $V$ because on $(0,T_Y-)$ $Y$ is uniformly zero. Thus, using the random horizon $T_Y$ allows focusing on the effect of $V$ on $Y$ (in our example, of blood pressure on dementia).

 The case of death is special. In short-term studies or in studies with young subjects, it may not be necessary to model death. In many studies, however, and especially in ageing studies, this is necessary. Death can be modelled by a $0-1$ counting process. It must be realized that all the other processes are defined for living subjects. Therefore, the maximum horizon for studying WCLI is $T_D$, the time of death. This is developed in Section \ref{ageing studies}.

\section{The dynamic approach to causal reasoning taking death into account}\label{ageing studies}
\subsection{The particular significance of death}\label{significanceofdeath}

In ageing studies, one of the most important events that we have to consider is death. This is why the illness-death model is important in such studies, but death should also be taken into account when studying a quantitative marker. The critical point is that  death is not an event which is on the same footing as other events that can happen to subjects.
 Even if the vital status is part of the state associated to each subject, this part of the state has a very special meaning, in that all the other components of the state are defined only for a living subject. The consequence is that causal influences must be defined on a maximum horizon $T_D$, where $T_D$ is the time of death. This has also a consequence for the graph representation of the system; we will represent influences of the components of the state on the death process, but not influences of the death process on other components. Since the death process is a special process, we may represent it by a special symbol, for instance a star: $\bigstar$.
 For  instance if we are interested in dementia, the state can be represented by a bivariate counting process $(D,Y)$, respectively counting dementia and death. However, dementia is defined only for a living subject: after death the subject does not exist anymore and cannot be qualified as demented or not demented. When we investigate the causal influence of a factor, we should obey to the following

\noindent  {\bf Rule 1} First look at causal influence on death, then on  influence on other processes.

\subsection{Example of causal reasoning in dementia}
\subsubsection{Possible systems}\label{pos-systems}
Let us look at systems including death and dementia. In the simplest system, there are two processes, dementia and death, so that the state process is a bivariate counting process $\bX'=(D,Y)$. This can also be represented by an illness-death process \citep{commenges2007likelihood}. With the Markov property, the illness-death process is specified by the transition intensities $\alpha_{01}(t)$, $\alpha_{02}(t)$ and $\alpha_{12}(t)$.
 The intensity of the dementia process, defined on $(0,T_D)$, is:
$$ \lambda_{Yt}=1_{\{Y_{t-}=0\}}\alpha_{12}(t),$$
and the intensity of the death process is:
$$\lambda_{Dt}=1_{\{D_{t-}=0\}}[1_{\{Y_{t-}=0\}}\alpha_{02}(t)+1_{\{Y_{t-}=1\}}\alpha_{12}(t)].$$
As noted in Section \ref{significanceofdeath}, influences are studied on $(0,T_D)$ so that the only possible causal influence is that of dementia on death.
Dementia influences death if the transition intensities $\alpha_{02}$ and $\alpha_{12}$ (resp. death rates for non-demented and demented) are different, and this is symbolized by $Y \dinf{\bX'} D$.
The graph reduces to two nodes (dementia and death) and one arrow from dementia to death. This simplified form of the graph allows representing influences in more complex models, which cannot be done with the conventional graphs for multistate models.

We may be interested in the effect of blood pressure on dementia.
This is particularly interesting  because we can consider blood pressure as a modifiable factor since there are anti-hypertensive treatments. So, the question of the possible causal influence of high blood pressure on dementia is of practical importance in public health.
 We cannot, however, dissociate the issue of causal influence of high blood pressure on death and on dementia. Anti-hypertensive treatments may also decrease the risk of death but we have also to consider the theoretical possibility that treating hypertension increases the risk of death. So the parameters of importance are both the effect on the intensity of death and the effect on the intensity of dementia.
In order to approach causal inference, we must introduce other important factors, generally considered as explanatory variables in a multistate model. In this framework, fixed variables are called ``attributes'', while internal time-dependent variables are components of the state process (there may also be external time-dependent variable, not treated in this paper). Attributes are linked to the identity of the subject, like gender or more generally genetic factors. It is important to distinguish attributes from state because attributes cannot be influenced. To distinguish them visually we will represent attributes by squares in the graphs.

\begin{figure}[h!]

\centering
\includegraphics[scale=0.55]{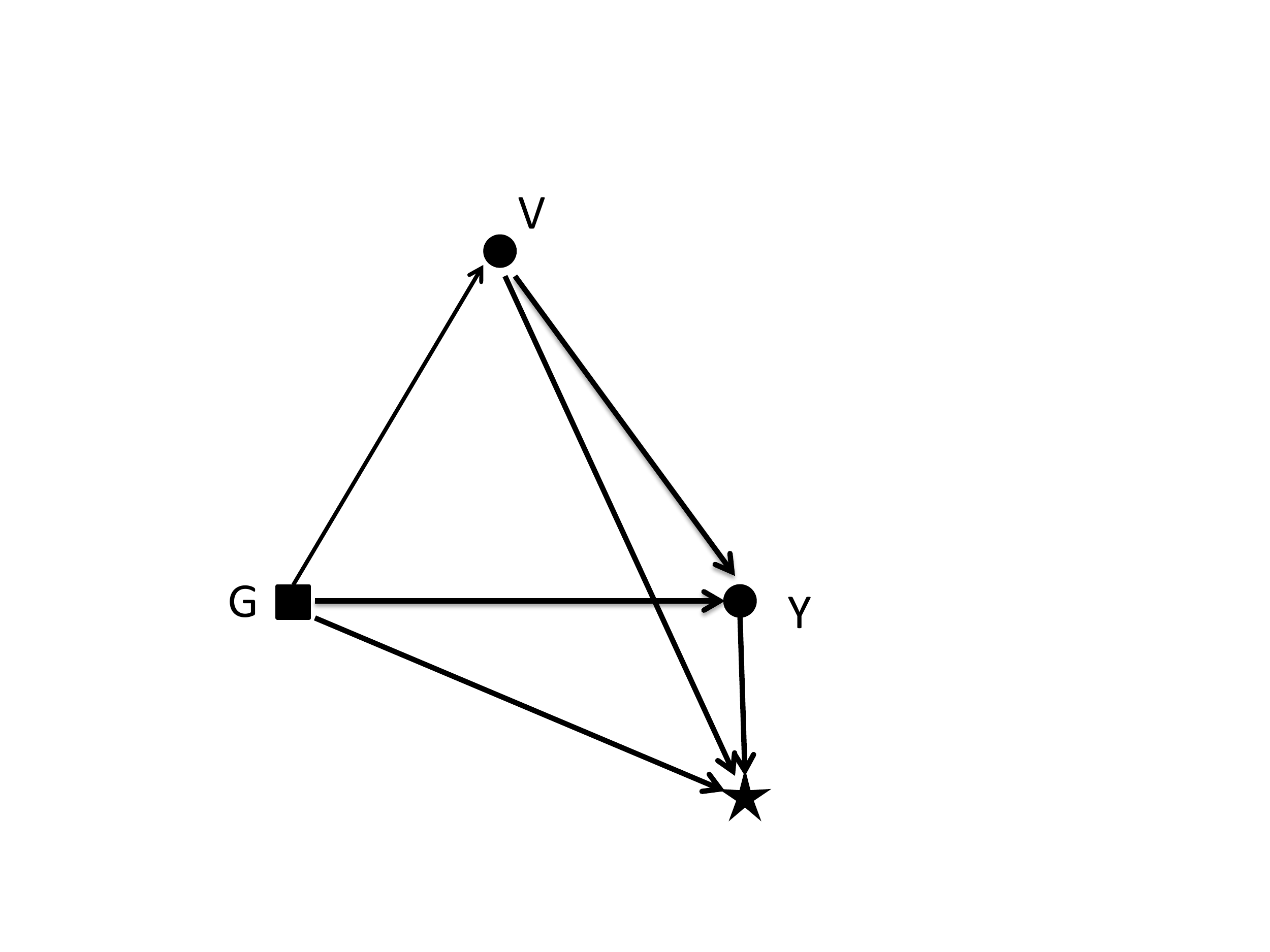}
\caption{Influence graph for dementia. $Y$ (dementia or cognitive ability) and $D$ (death, represented by a star) are the processes of interest, $V$ represents a modifiable factor (blood pressure) and $G$ represents the attributes (gender, genetic factors). \label{graphDementia}}
\end{figure}

\begin{figure}[h!]
\centering
\includegraphics[scale=0.55]{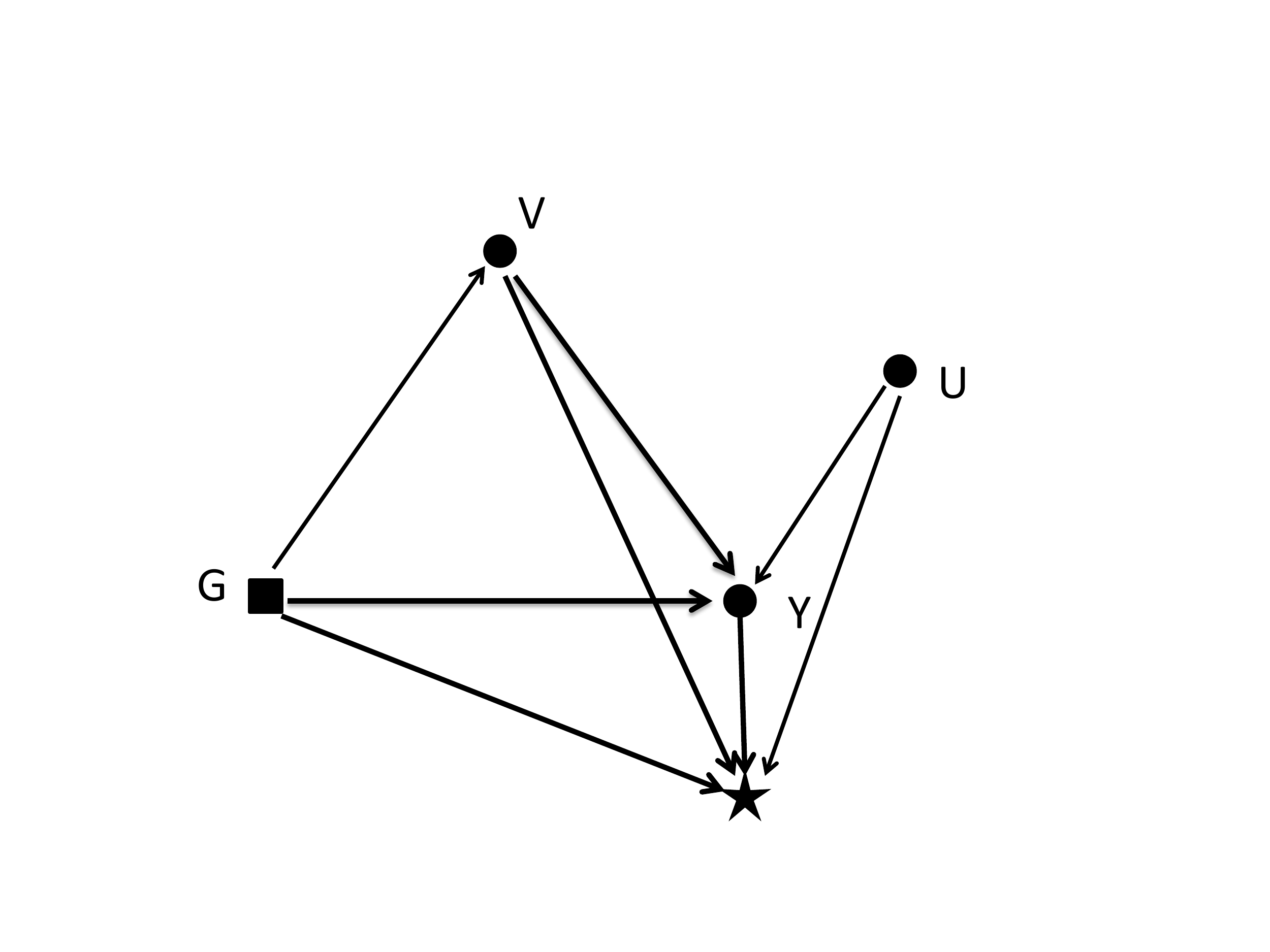}
\caption{Influence graph for a physiological process $Y$ and $D$ (death, represented by a star); $V$ represents a modifiable factor (blood pressure) and $G$ represents the attributes (gender, genetic factors); $U$ is another process influencing $D$ and $Y$ but not $V$. \label{graphPerfectSystem}}
\end{figure}

So, we may consider the system $\bX=(D,Y,V,G)$, where $V$ represents the blood pressure process and $G$ the attributes. This system is depicted in Figure \ref{graphDementia}.
We will assume that $\bX$ is a perfect system, or more realistically a NUC system for $(V,D)$ and for $(V,Y)$ (see Section \ref{causalinfluences}). If we can find $U$ which influences $Y$ or $D$, $\bX$ is not perfect; if $U$ does not influence $V$
 then, $\bX$ is still a NUC system for $(V,D)$ and $(V,Y)$; the graph of a system including such $U$ is represented in Figure \ref{graphPerfectSystem}.
For instance if $G$ does not include educational level, the system is not perfect for $Y$ because it has been shown that educational level influences dementia; if educational level does not influence blood pressure, the system is still be a NUC system for $(V,Y)$.

\subsubsection{Preferable order}
In view of the situation described in Section \ref{pos-systems} the choice between two values of the possibly manipulable factor $V$ is not always obvious. The aim of this section is to formalize the cases where it is. 

Assume that we know the law of $\bX$; then we can compute  the probability of being alive non-demented, alive demented and dead for any time (age) $t$ for any given value of $V=v$. If the probability of  being dead and the probability of  being demented are both lower for value $v_1$ compared to value $v_2$, $v_1$ is clearly preferable to $v_2$. The computation can be done for given $G=g$ or marginally to $G$.
\begin{Definition}[{\em Preferable} partial order] \label{preferable}
Let $\bX=(D,Y,V,G)$ a NUC system for $(V,D)$ and $(V,Y)$ and denote by $\PP^{*v_1}$ the true probability law when the value of $V$ is $v_1$, and $\Ee^*_{v_1}$ the expectation under this true probability: $v_1$ is {\em preferable} to $v_2$ if $\PP^{*v_1}(D_t=1) \le \PP^{*v_2}(D_t=1)$ and $\Ee^*_{v_1}(Y_t) \le \Ee^*_{v_2}(Y_t)$, with strict inequality holding for some $t$. \end{Definition}

If $Y$ is a $0-1$ process, $\Ee^*_{v_1}(Y_t)=\PP^{*v_1}(Y_t=1)$; the definition also applies to a quantitative $Y$ when high values of $Y_t$ are detrimental.

To fix the ideas, let us assume that the true law of $Y$ and $D$ given $G$, is specified by the intensities:
\begin{eqnarray}
 \lambda_{Yt}&=&1_{\{Y_{t-}=0\}}\alpha_{Y}(t), \mbox{where}~~ \alpha_{Y}(t)=\alpha^*_{0Y}(t)e^{\beta^*_1 G+\beta^*_2 V_t}\\
 \lambda_{Dt}&=&1_{\{D_{t-}=0\}}\alpha_{D}(t), \mbox{where}~~ \alpha_{D}(t)=\alpha^*_{0D}(t)e^{\gamma^*_1 G+\gamma^*_2 V_t+\gamma^*_3Y_t}
\end{eqnarray}
The intensity\index{intensity} for $Y$ is defined only on $[0,T_D]$. These equations describe the true law of our processes and not a model (that is a family of laws), and this is the meaning of putting a ``*'' in superscript of the symbols; this is to make clear that we are not treating here the inference problem. We have a perfect or NUC system and we know the true law; in this ideal situation what will we do?

If both $\beta^*_2$ and $\gamma^*_2$ are positive, a sufficient condition for ``$v_1$  preferable to $v_2$''  is that $v_1(t)\le v_2(t)$ for all $t$, with strict inequality for some $t$.
In this case, we could deduce that lowering blood pressure from $v_2(t)$ to $v_1(t)$ would lead to a preferable situation.

As a second example, we consider the case where the process of interest $Y$ is the global cognitive ability.
The true law could be that $Y$ is a diffusion process: $\dd Y_t=\lambda_{Yt}\dd t + \dd B_t$, where $B$ is a Brownian motion.

Let us assume that the true law given $G$, is specified by the intensities:

\begin{eqnarray}\label{example2}
 \lambda_{Yt}&=&\beta^*_0(t)+\beta^*_1 G+\beta^*_2 V_t \label{eq:dynY}\\
 \lambda_{Dt}&=&1_{\{D_{t-}=0\}}\alpha^*_{0D}(t)e^{\gamma^*_1 G+\gamma^*_2 V_t+\gamma^*_3Y_t}\label{eq:dynD},
\end{eqnarray}
where $\beta^*_0(\cdot)$ and $\alpha^*_{0D}(\cdot)$ are baseline functions.
As before, $\lambda_{Yt}$ is defined only on $[0,T_D]$. See the link between this modeling and the standard mixed-effect modeling for $Y$ in Section \ref{mixed-effect}.
Here, $Y$ (cognitive ability) is clearly a construct and cannot be observed in continuous time. It is indirectly measured by cognitive tests, necessarily at discrete times. If the system is perfect or NUC, we can make computation that can be used for choosing the best value of $V$, in the case where $V$ can be manipulated. For each value $V=v$, we can compute the probability of being alive at any $t$, then given alive at $t$, the distribution of $Y_t$. If, for all $t$, the probability of being dead is lower and the mean value of the cognitive ability is higher for value $v_1$ compared to value $v_2$, $v_1$ is clearly preferable to $v_2$.

When $v_1$ and $v_2$ cannot be ordered this way, a utility function $\Phi(v)$ has to be constructed.  If $V$ can be manipulated, one can choose the trajectory $v$ which maximizes $\Phi(v)$.

\subsubsection{Conditional and marginal effects}
In the ideal situation of a perfect system, the compensators of the processes of interest, $D$ and $Y$, encapsulate the effect $V$, and these effects can be summarized by some contrast between the compensators of $D$ and $Y$, respectively, for two different values of $V$.

Let us assume that $\bX=(D,Y,V,G)$ is perfect. The causal effect of $V$ can be summarized by any contrast of the compensators for two values of $V$, say $v_2$ and $v_1$: the simplest is the difference or the ratios of the intensities: if we use the difference for $\lambda_{Yt}$ in Equation (\ref{eq:dynY}) we find $\beta^*_2 (v_{2t}-v_{1t})$, and if we use the ratio for  $\lambda_{Dt}$ in Equation (\ref{eq:dynD}) we find $e^{\gamma^*_2 (v_{2t}-v_{1t})}$, the hazard ratio. Other interesting contrasts bear on the expectation of the processes; they are in fact functions of the compensators, so that contrasts between expectations can also be considered as contrasts between compensators, with, however, a direct interpretation. So we may contrast $\PP(D_t=0|G,V=v)$ and $\Ee(Y_t|G,V=v)$  for different values of $v$.

However, we may be interested in marginal effects, for two reasons. First, we may wish to know what is the global effect of $V$ in a population. In our example, $G$ has a distribution in the population, so we could be interested in the marginal effect with respect to $G$. Second, we may doubt that the system we have built is a perfect system; it may more realistic to assume that it is a NUC system for $(V,D)$ and $V,Y)$.
There are two cases. First,  if $G \wli{\bX} V$, then  $\bX'=(D,Y,V)$ is still NUC for $(V,D)$ and $(V,Y)$. In that case we could look at contrasts between the compensators of $D$ and $Y$ in this imperfect system. However, even in that case, it is not likely to give very insightful interpretation. One example, in the simpler case of a survival model with frailty (here $G$) has been given by \citet{aalen2008survival}: in the case of a proportional hazard for $V$, the marginal hazard ratio is no longer proportional and in some cases of the distribution of the frailty the log marginal hazard ratio can have a sign opposite to the log conditional hazard ratio; see also \citet{roysland2015does}.
So, when looking at marginal effects, it is better to use as criteria teh marginal (or partly marginal) expectations of the processes of interest:
$\PP(D_t=0|V=v)$ and $\Ee(Y_t|V=v)$ (rather than with dynamic parameters such as hazard ratios).

In the second case $G \dinf{\bX} V$; then it is a confounding factor so that we cannot use $\bX'$ for finding the marginal causal effect of $V$.
However, we can still compute it from the law of $\bX$.

\section{Observation and inference}\label{sec:observation}
\subsection{Generalities}
In the stochastic system approach, there is a clear distinction between the physical system and the observations that we make.
Observing the processes in continuous time until $T_D$, we have complete information. It often happens that we have incomplete observations (the observed sigma-field is not included in the sigma-field generated by $\bX$).
If the mechanism leading to incomplete data (mlid) is ignorable (see Section \ref{section:RIP}), and assuming a well specified model and identifiability, we can consistently estimate the parameters from observations by maximum likelihood or a Bayesian approach.

In our first example where $Y$ is dementia, if all attributes and processes are observed in continuous time, with possible right-censoring, maximum likelihood estimation can easily be done by splitting the problem into several conventional survival problems, using the technical trick explained in \cite{Andersen2002}. This will not be possible either if some attributes which have an influence on both dementia and death are not observed, or if $V$ or $Y$ are observed in discrete time (inducing interval-censoring). In both cases one can write the likelihood conditional on the complete data, and obtain the observed likelihood by taking the expectation. This leads to the computations of numerical integrals.

If $Y$ is a quantitative phenomenon, we must acknowledge that we never exactly observe it. There is always an observation error. This is also true for a binary state where observation errors are called ``misclassification errors''; for instance there may be errors in the  diagnosis of dementia. However, in the binary case, it may be acceptable to neglect the observation error. In our example, we identify ``dementia'' and ``diagnosis of dementia''. It is less acceptable to identify a score to a psychometric test and cognitive ability; so we must represent a noisy observation.

\subsection{The response indicator processes}\label{section:RIP}
The statistical question is to estimate the compensators; for this, we need observations and models. We will assume that we have well-specified models (that is, the true law is within the model). However, we rarely have complete observations of the system over the  horizon of interest.

As in \citet{commenges2005likelihood} and \citet{commenges2007likelihood}, for each physical process,  we introduce a response indicator process (RIP) which takes value $1$ if the physical process is observed at time $t$, zero otherwise. This representation (called ``Time coarsening for stochastic processes'' (TCMP) ) allows us to represent very general mechanisms of censoring or coarsening. For instance in case where death is observed with right-censoring, the RIP $R_D$ is : $R_{Dt}=1_{t \le C}$, where $C$ is the conventional censoring variable. In case where $Y$ is observed at discrete times $t_1, \ldots, t_m$, the corresponding RIP is $R_{Yt}=1$ if $t\in \{t_1,\ldots,t_m\}$, and $R_{Yt}=0$ otherwise. In the case where the RIPs are fixed  and assuming we have a well-specified model for $X$, it is possible to write the likelihood and to estimate the true law by maximum likelihood.

When the RIPs are random, the observation can be represented by $(R, {\bR}\bX)$, where $R$ is the vector of RIPs and where {\bR} is a diagonal matrix with diagonal given by $R$.  The main issue is whether the mlid is ignorable, that is whether we can estimate the true law by maximizing a ``partial likelihood'' which does not include the likelihood terms coming from the observation of the RIPs themselves. Extensions of the concepts of MCAR, MAR, et MNAR have been developed. The concept of coarsening at random (CAR) for random variable was introduced by \citet{Heitjan1991} as an extension to the missing at random (MAR) concept. This was extended to a more general context by \citet{gill1997coarsening}. Here we shall use the definition of CAR for processes proposed by \citet{commenges2005likelihood} and \citet{Commenges2007}. It is obvious that if the RIPs are independent from $\bX$ the mechanism is ignorable. The important concept is that of coarsening at random (CAR) for processes, denoted ``CAR(TCMP)'', where TCMP stands for ``time coarsening model for processes''. When $Y$ has a continuous state-space, we must add a model for observation errors; it is also possible to add a model for misclassification when $Y$ has a discrete state-space.



\subsection{Coarsening at random for processes}
  We use here a dynamical version of CAR(TCMP) called ``CAR(DYN)''.
 When the condition called CAR(DYN) holds, the mlid is ignorable and estimation can be done by maximizing the ``partial'' likelihood, that is writing the likelihood as if the observation times, or more generally the RIPs, were fixed.
Essentially CAR(DYN) holds if the law of $R$ conditional on $\bX$ depends only on the past observed values of $X$. Here is a rapid definition.

 Assume that we can represent $(R_t)$ by a point process $(N_t)$. Denote by $(\N_t)$ the filtration generated by the process $(N_t)$.  We define the filtration $(\Ob_t)$ as the family of $\sigma$-fields
$\Ob_t=\sigma(N_u,R_u\bX_u, 0\le u \le t)$. We define also the filtration generated by $\bX$, $(\X_t)$, and we denote by $\X$ the sigma-field generated by $\bX$ on the maximum horizon. Let us call
$\Lambda^{\Ob,N}=(\Lambda^{\Ob,N}_t)$  and
$\Lambda^{\FF^*,N}=(\Lambda^{\FF^*,N}_t)$ the compensators of $N$
in the filtrations $(\Ob_t)$ and $(\FF^*_t)$ respectively  where $(\FF^*_t)$ is the family of
$\sigma$-fields $\FF^*_t= \X \vee \Ob_t, t\ge 0$.

\begin{Definition}[CAR(DYN)]

We will denote CAR(DYN) the condition:

 Under the probability laws that we use : $(\Lambda^{\Ob,N}_t) =
(\Lambda^{\FF^*,N}_t)$, (up to indistinguishability).
\end{Definition}

Intuitively this says that the dynamics of the RIPs only depends on the past observed values of the system $\bX$ and of the RIPs themselves; it depends neither on future nor on unobserved past values of $\bX$.  In the case of a multistate models, \citet{gruger1991validity} call ``doctor's care'' the case where observation times are decided as a function of the observed state of the patient. In that case the RIP is clearly not independent of the state of the patient but the observation process is CAR(DYN) and thus, is ignorable.

\subsection{The case of random effects and pure latent processes}\label{sec:random effect}
Random effects are often included, especially in models for quantitative $Y$. Introducing random effects $U$ is a way to represent an attribute in the system  but which is unobserved, that is, with null RIP $R_{U}=0$. It is called ``random'' because for inference we cannot condition on the observation of $U$, as is done for observed explanatory variables or processes.

A latent process can be considered as a time-varying random effect. Such a ``pure'' latent process can be represented as a process in the system, say $U$, having an identically null RIP: $R_{Ut}=0$, for all $t$.

In spite of this lack of observation we can still estimate the effect of the unobserved attribute $U$ on $Y$ and $D$, and the conditional effect of $V$  (by using the likelihood, which is marginal with respect to the random effects), subject to identifiability:
in particular, it is not possible to identify an effect on both $V$ and $Y$, that is to remove the potential confounding effect of an unobserved attribute. Thus, we have to assume that $U \wli{\bX} V$.

Adding a random effect or a pure latent process to a system  $\bf X$ assumed to be NUC is an attempt build a perfect system.
Suppose that attributes $U_1,\ldots, U_k$ influence $Y$ and $D$, and that the system $(D,Y,V,G,U_1,\ldots,U_k)$ is perfect; assume their influence on $Y$ is through a linear form $U_Y=\beta^*_{Y1}U_1+\ldots +\beta^*_{Yk}U_k$, and that on $D$ is through $U_D=\beta^*_{D1}U_1+\ldots +\beta^*_{Dk}U_k$.
If we do not (or cannot) want to distinguish the effects of the $U_j$s, it is equivalent to work with  $(D,Y,V,G,U_Y,U_D)$ that we can consider as perfect for $(D,Y)$. We may also simplify, assuming the same form $U$ influences both $Y$ and $D$ leading to the system $\bX^{\tilde M}=(D,Y,V,G,U)$ (meaning by this notation that $X^{\tilde M}$ is close to a perfect system $\bX^M$.

\subsection{Latent processes and more general observation equations}\label{sec:latent process}
 There is, another case of latent process which is a process which can be indirectly observed. For instance, ``cognitive ability'' can be considered as a latent process that can be indirectly observed by psychometric scores and even, by diagnosis of dementia (see Section \ref{sec:Illustration}).

This leads us to present a more general observation equation which unifies the observation of events and quantitative processes, whether latent or not. In its general form it also includes observations schemes used in mechanistic models \citep{prague2012treatment}. In mechanistic models we may observe combinations of different components of $\bX$ (for instance we observe only the sum of infected and non-infected CD4+ T lymphocytes). Thus the observable is $g(\bX_t)=(g_1(\bX_t),\ldots,g_k(\bX_t))$; the observable process is measured at inspection times and the observation is noisy. In more standard statistical applications one can often separate the observations of the different processes of the system. For sake of  simplicity we give the general scheme of observation in the latter case for a process $Y$ of $\bX$.  The general form of the observation is $(R_{Yt},Z_t, 0<t<\tau )$, where $Z_t$ is the observed process; if the mlid is ignorable we do not need to model the distribution of $R_{Yt}$. $Z_t$ is obtained using potentially three ingredients: the transformation function $g$, the RIP and the noise. We describe a first model, ``Model-a'', involving three stages and in which the noise is a measurement error:
\begin{enumerate}
\item Potential observable: $g(Y_t), 0<t<\tau $;
\item Measured: $R_{Y_t}g(Y_t), 0<t<\tau $;
\item Noisy observation: $Z_t=h(R_{Y_t}g(Y_t),\varepsilon_{Yt}), 0<t<\tau $;
\end{enumerate}

Completely non-observed latent processes are characterized by an identically null RIP ($R_{Y_t}=0$). One still qualifies as ``latent'' a process which is indirectly observed, that is with unknown $g(.)$, often not one-to-one, and which is measured at discrete time. We then need a model for $g(.)$ involving specific parameters that we will have to estimate \citep{proust_bcs2006}.
We do not call ``latent'' a process which can be observed ``directly'', that is with $g(.)$ the identity function or a known one-to-one function.

The stage 2 of Model-a is related to the times of measurement and is characterized by the RIP (see Section \ref{section:RIP}) .
Stage 3 is the possibly added noise; we assume in general that the $\varepsilon_t$'s are independent from $Y$. Often, processes with binary state-space are considered to be observed without noise; see for instance the statistical analysis of dementia proposed by \citet{joly2002penalized}. However, there may be misclassification errors. In that case we observe a binary variable with a Bernoulli conditional distribution specified by $\PP(Z_t=1|Y=1)$ and $\PP(Z_t=0|Y=0)$ (called in another context ``sensitivity'' and ``specificity''). This can be represented by the function $h(R_{Y_t}g(Y_t),\varepsilon_{Yt})=Y_t 1_{\varepsilon_t<c_1}+(1-Y_t)1_{\varepsilon_t<c_2}$.
For continuous state-space non-latent processes, the observation is generally in discrete time $t_j, j=1,\dots,m$; an additive error model is often used:
$h(g(Y_{t_j}),\varepsilon_{Yt_j})=Y_{t_j}+\varepsilon_{Yj}$.

Another model, ``Model-b'',  for introducing noise is possible; here the observable is a noisy version of $Y$ and it involves two stages:
\begin{enumerate}
\item Potential observable: $g(Y_t,\varepsilon_t), 0<t<\tau $;
\item Observed: $Z_t=R_{Y_t}g(Y_t,\varepsilon_t), 0<t<\tau $;
\end{enumerate}
If $g$ is  a linear function, the two models are identical. \citet{prague2012treatment} used  Model-a for observation of a their mechanistic system while \citet{proust_bcs2006} as well as \citet{ganiayre2008latent} used Model-b for observation of a latent trait representing cognitive ability; see Section \ref{sec:Illustration}. While it is theoretically possible to mix the two models, this would not be practically identifiable in most cases.

The concept of CAR(DYN) can be extended to these more general models of observation. The definition is unchanged and $\Ob_t$ still represent the observed sigma-field, but it must be defined in terms of $Z$:
$$\Ob_t=\sigma(N_u,Z_u, 0\le u \le t).$$

\section{Estimating the effect of a factor on a physiological state in presence of death}\label{effect-on-Y}
\subsection{General setting}
We now tackle in some detail the issue of estimating the effect of a factor $V$ on a physiological process $Y$ in the system $\bX=(D,Y,V,G)$ assumed  NUC system for $(V,D)$ and $(V,Y)$, or in a system enlarged to an unobserved process $U$, $\bX^{\tilde M}=(D,Y,V,G,U)$ that we hope to be close to a perfect system for $(D,Y)$ (see Section \ref{sec:random effect}).

In system $\bX^{\tilde M}$, the way $V$ is involved in the compensators of $D$ and $Y$ in the true probability {\em is} (by definition) the effect of $V$, conditional on all other factors; in $\bX$, it is marginal on $U$. It is then possible to summarize these effects by suitably chosen contrasts between the values of the compensator for different values of $V$. Recall that this compensator is only defined on $(0,T_D)$ and that we have first to look at the effect of $V$ on death. Then if we know these two compensators, we can look at preferable values for $V$ if we can manipulate it.

In practice we have to estimate the compensators thanks to observations, as described in Section \ref{sec:observation}.
In this section, assuming that $G$ and $V$ are completely observed, we examine whether the mlid is ignorable (using the CAR(DYN) condition) when $R_Y$ depends on $Y$, in both situations where $Y$ is a 0-1 process and where $Y$ is a quantitative process. The conclusions are summarized in Table \ref{Table:Typology}.

\subsection{Typology of the cases where CAR(DYN) holds in a NUC system}

\subsubsection{Typical observation when $Y$ is a $0-1$ counting process}
 We assume that death is observed in continuous time with right-censoring. The most conventional case is that $Y$ is also observed with right-censoring. It is however often more realistic to acknowledge that $Y$ is observed in discrete-time (leading to interval-censored observations).
Let us look at the CAR(DYN) condition for $R_Y$. First, if $R_Y$ is completely independent of $\bX$, CAR(DYN) holds. $R_Y$ may be influenced by $G$ and $C$ but this is not a problem if we assume they are completely observed (which we do in this section). Secondly, if $R_Y$ is influenced by an unobserved process $U$ which also influences $Y$, CAR(DYN) cannot hold.

\noindent{\bf Case where $R_Y$ may be influenced by $Y$.} We must distinguish between the case with right-censoring and the case with interval-censoring. If $Y$ is right-censored it is clear that CAR(DYN) holds (because right-censoring produces an observation in continuous time until censoring, so all past values of $Y$ have been observed at the time of censoring).
If $Y$ is interval-censored, then the law of the RIP may depend on unobserved values of $Y$. Thus, we do not have necessarily CAR(DYN). For instance CAR(DYN) holds if $Y$ is observed at fixed visit times which may have been planned in a cohort study. However, there may also be loss to follow-up which may depend on unobserved values of $Y$.


As for the RIP of $D$, $R_D$, the same conclusions holds as for $R_Y$. However, since most often $D$ is observed in continuous time until censoring CAR(DYN) generally holds.

{\bf Example 1:} suppose that observation of vital status ($D$) is done until a fixed date, (administrative censoring) or, by design, until one year after $\tilde T_Y$, the time at which $Y$ has been first observed to have jumped; then, $R_D=R^a_D1_{\{t<\tilde T_Y+1\}}$. 
 CAR(DYN) holds in this case.

{\bf Example 2:} $R_D$ may be influenced by unobserved values of $Y$, in which case CAR(DYN) does not hold: this may happen if $Y$ is observed in discrete time and the probability of loss to follow-up depends on $Y$ (for instance, demented people may enter into institution or refuse to participate to the study). However, this should be the case only in a badly designed study because in many countries there exist death registers, so that censoring of vital status can be reduced to administrative censoring.

\subsubsection{Typical observation when $Y$ is a quantitative process}
Here, the most conventional case is that $Y$ is observed in discrete-time. Essentially the same conclusions can be driven as for the case where $Y$ is a $0-1$ process observed in discrete-time. If there are observation errors, we observe for instance, under Model-a (see Section \ref{sec:latent process}) $Z_j=Y_{t_j}+\varepsilon_j$, for the $t_j$s such as  $R_{Yt_j}=1$; note that we have necessarily $t_j<T_D$ since $Y$ is not defined after $T_D$. We generally make the assumption that the $\varepsilon_j$s and the $Y_{t_j}$s are independent. If $R_Y$ were only influenced by observed values of $Y$ then CAR(DYN)  holds (the doctor's care scheme of \citet{gruger1991validity}). An example with a quantitative process would be that $Y$ represent the concentration of T-CD4+ lymphocytes, and the doctor determines the next visit for an HIV infected patient as a function of the observed CD4 counts (the $Z_j$s).
However, it may happen that $R_Y$ is influenced by unobserved values of $Y$, in which case CAR(DYN) does not hold; note that $Y$ is never exactly observed because of the observation error. For instance opportunistic diseases may be influenced by the true value of CD4+ T-lymphocytes concentration and occurrence of such a disease may precipitate a new visit to the doctor.

\begin{table}
\caption{Cases where CAR(DYN) holds (that is, the mlid is ignorable) when $Y$ influences $R_Y$, according to whether $Y$ is $0-1$ or quantitative and to the continuous or discrete-time observation scheme. A degree of plausibility is indicated for each situation.}\label{Table:Typology}
\begin{center}
\begin{tabular}{|l|l|l|}
\hline
  & Continuous time & Discrete-time \\
\hline
& & \\
$Y$ : $0-1$ & plausible &  realistic \\
& & \\
 & CAR(DYN) holds & CAR(DYN) does not hold   \\
 \hline
 & & \\
$Y$ : quantitative & not plausible  & nearly always the case \\
& & \\
&  CAR(DYN) holds & Doctor's care: CAR(DYN) holds\\
&   & otherwise: CAR(DYN) does not hold  \\

\hline

\end{tabular}
\end{center}
\end{table}

\section{Estimation with the incomplete system not including Death}\label{sec:incompletesystem}
\subsection{Treating death as a drop-out}
\subsubsection{Treating death as a drop-out: generalities}
Often death is treated as a drop-out. What happens if death is ignored? Ignoring death means that we work with a smaller system $\bX'=(V,Y,G)$. If death has a non-negligible intensity this system is not correct; death has then to be treated as censoring, which is not correct because this is treating as a part of the observation mechanism what is in fact an important part (arguably the most important part) of the physical system. The first problem is that when examining the effect of a factor $V$, we should first look at its effect on death. This could be done in a separate analysis using the system $\bX''=(D,V,G)$, which is a correct system allowing us to estimate the marginal (wrt Y) effect of $V$ on death, at the condition that $Y \wli {\bf X} V$.  However, for estimating the effect of $V$ on $Y$, the problem is that death considered as a censoring may be ``informative''.
\subsubsection{The case of drop-out}
Since in this approach death is treated as a drop-out, it is important to first study the case of possibly informative drop-out, and then examine the difference between drop-out and death. We examine the case where observation of $Y$ is made until drop-out. The drop-out process $S$ (S for the French word ``sortie'') is a $0-1$-counting process; we have $R_{Yt}=1_{S_t=0}$, $S$. If really a drop-out and not death, $S$ not part of the physical system, but rather of the observation mechanism. So we can consider the quantities of interest as being the conditional and marginal expectations $\Ee(Y_t|V,G)$ and $\Ee(Y_t|V)$, respectively. It is not interesting to consider $\Ee(Y_t|V,G,S)$ because $S$ does not belong to the physical system; nevertheless, this quantity exists and is equal to $\Ee(Y_t|V,G)$ since $S$ does not influence the physical system. Thanks to our approach separating system and observation, we can get here a clear result!

\begin{figure}[h!]

\centering
\includegraphics[scale=0.55]{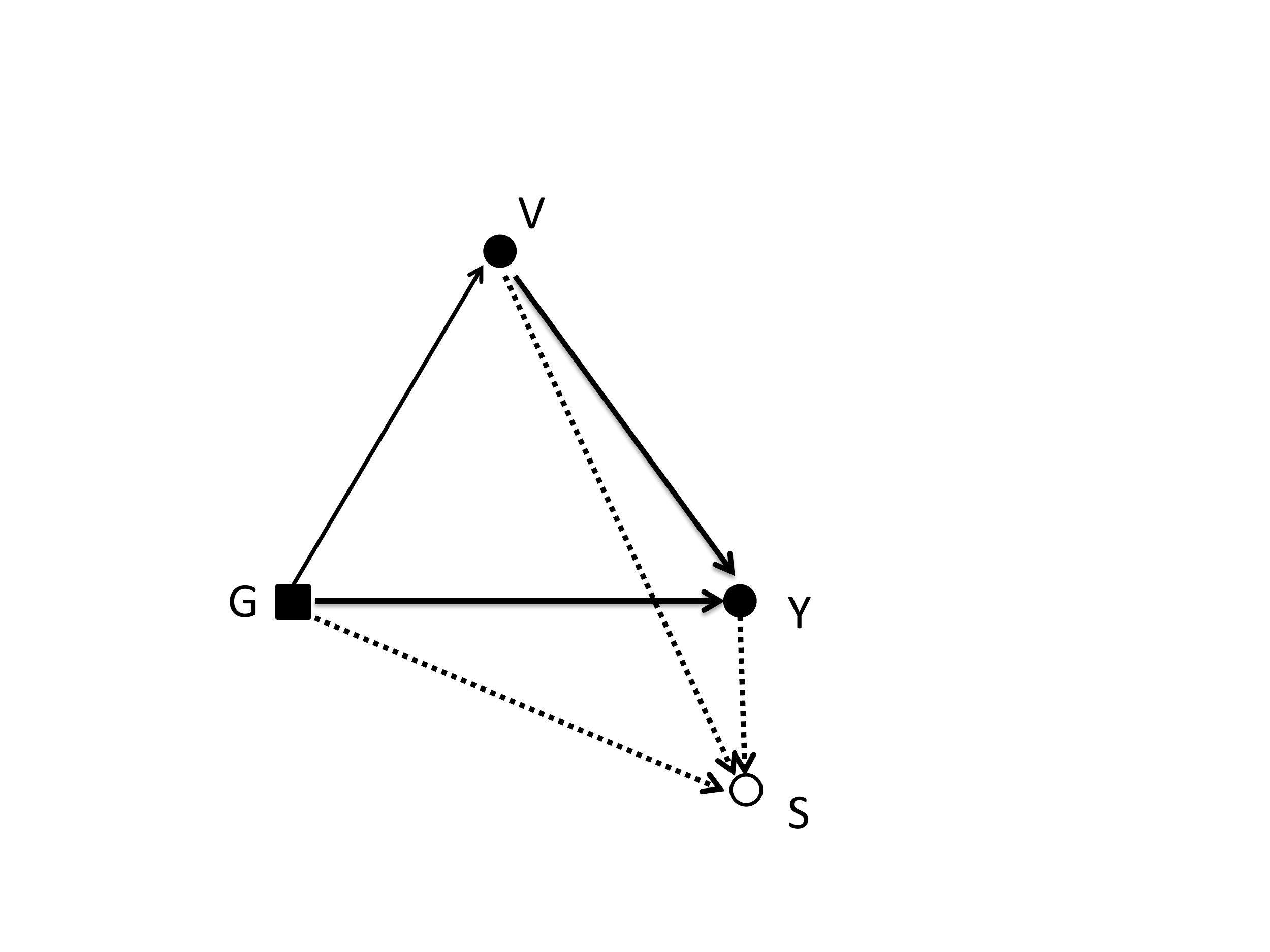}
\caption{Graph of the extended system including the drop-out process; this process is symbolized by an open circle and the arrows toward it are in dotted lines to distinguish it from the physical system.\label{graphDropoutMNAR}}
\end{figure}

\begin{figure}[h!]

\centering
\includegraphics[scale=0.55]{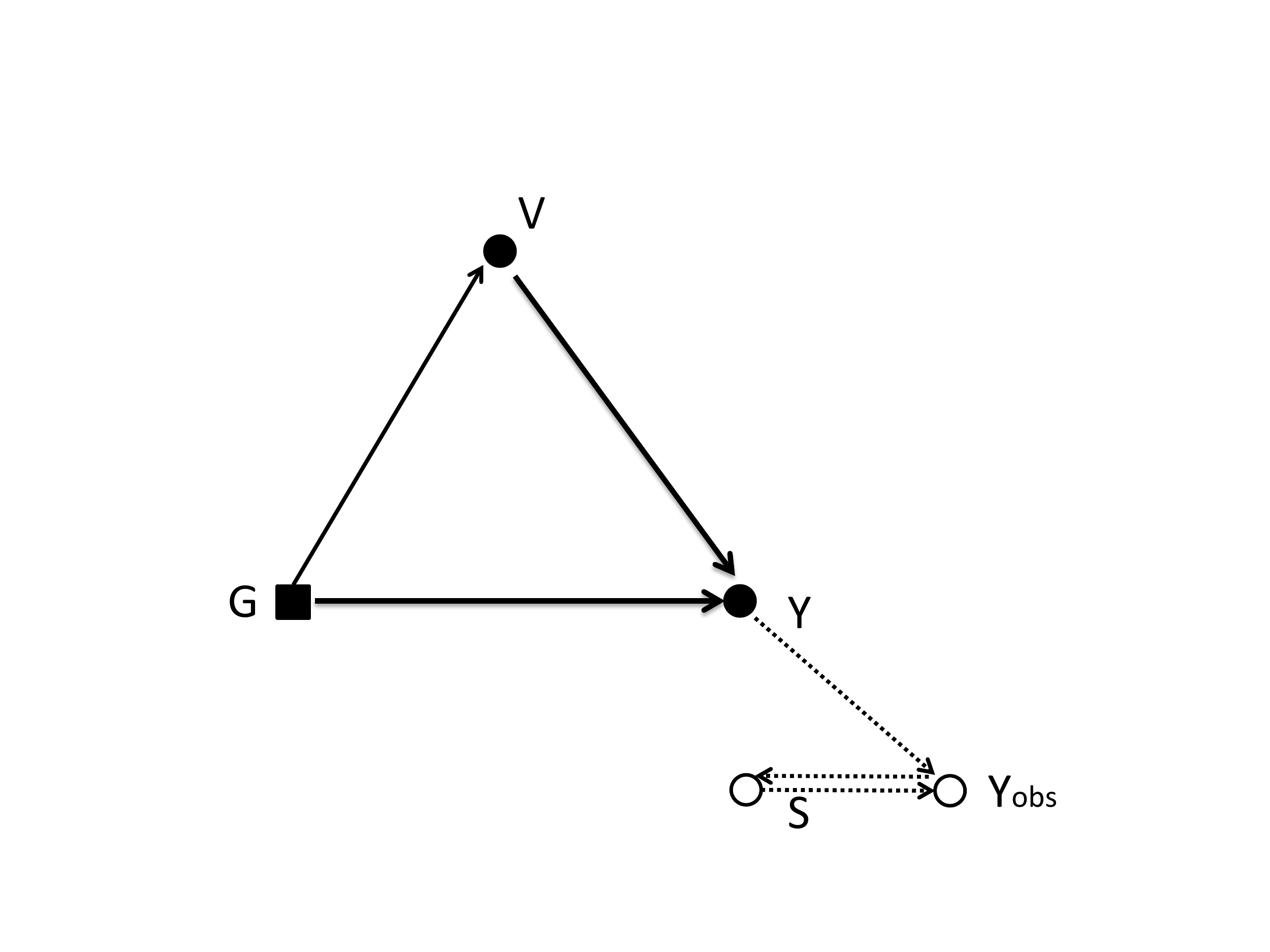}
\caption{Graph of the extended system including the drop-out process when the mlid is ignorable; this process is symbolized by an open circle and the arrows toward it are in dotted lines to distinguish it from the physical system; in that case it is influenced by $Y$ only through the observed part of $Y$ denoted $Y_{obs}$.\label{graphDropoutMAR}}
\end{figure}

The question for inference is whether the mlid is ignorable or not. As we have already said, in the CAR(DYN) case the mlid is ignorable; otherwise we have to model $S$. It is as though we included $S$ in an extended system including the physical system and processes belonging to the observation mechanism. We can represent graphically this extended system, with the convention that $S$ is represented by an open circle because it does not belong to the physical system.
 Figure \ref{graphDropoutMNAR} represents the graph of the extended system while  Figure \ref{graphDropoutMAR} represents the graph of the extended system when the mlid can be ignored.

\subsubsection{Death as drop-out ?}
In presence of death, when we wish to describe the situation at time $t$ we must first look at $D_t$, and if $D_t=0$ we can look at $Y_t$. In terms of conditional expectation, we have first to look at $\PP(D_t=0|G,V)$ and then at $\Ee(Y_t|G,V)$ on $D_t=0$, or in terms of marginal expectation at $\PP(D_t=0|V)$ and $\Ee(Y_t|V)$ on $D_t=0$; the expectations of $Y_t$ are implicitly conditional on $D_t=0$, although this is not really a conditioning but rather a question of definition: $Y_t$ is defined only if $D_t=0$. Thus, $\Ee(Y_t|G,V)$ is itself defined on $(0,T_D)$; it is not false to write $\Ee(Y_t|G,V,D_t=0)$ but since it is false to write $\Ee(Y_t|G,V,D_t=1)$, it is also false to write $\Ee(Y_t|G,V)$ without specifying that $D_t=0$.
\subsection{Typical observation when $Y$ is a $0-1$ counting process}
We must distinguish the continuous-time and discrete-time observation schemes.
If $Y$ is only right-censored, then CAR(DYN) still holds if censoring comes only from death. This has also been shown in \citet{andersen93}. Of course, deviation from CAR can come from other reasons of censoring. One could model the RIP as
$$R_Y=(1-D)R^a_Y R'_Y,$$
 potential problems coming from $R'_Y$ only.

If $Y$ is interval-censored, then in general CAR(DYN) does not hold, even if the visit times do not depend on $Y$. This is because generally $Y \dinf {\bX} D$ so that $R_Y$ depends on non-observed values of $Y$.

\subsection{Typical observation when $Y$ is a quantitative process}
\subsubsection{Theoretical analysis}
The most conventional case is that $Y$ is observed in discrete-time. We can use the same model for $R_Y$: $R_Y=(1-D)R^a_Y R'_Y,$.
Here $R'_Y$ is null everywhere except at observation times $(t_1,\ldots,t_m)$. If $Y \wli {\bX} D$ and $U \wli {\bX} D$ (where $U$ is a random effect), then CAR(DYN) holds. CAR(DYN) also holds if the vital status $D$ is influenced only through previously observed values of $Y$.
In this case, Death can be treated as a non-informative drop-out which simplifies the analysis. A funny example is that $Y$ represents the true weight of  a chicken and it is killed a random time after it has been {\em observed} to weigh more than 2kg; such a situation is not likely to happen in human health, although it may have applications in agricultural science.

However, if $Y \dinf {\bX} D$ or $U \dinf {\bX} D$, then CAR(DYN) does not hold.  The latter case will most of the time happen in epidemiology, for example if $Y$ represents psychometric tests, or a latent process indirectly measured by psychometric tests. So the question is to know whether the bias induced in that case can be large and in which cases is it large, in which cases is it negligible. Intuitively, the more accurately we can predict the unobserved values of $Y$, the smaller the bias. This accuracy depends on the frequency of the observation times and on the quality of the model. Of course the bias will also depend on the effect of $Y$ on $D$.
In fact the development for the observation model should imply not only the RIP but also a noise model. It is clear that the rate of death for instance could depend on the psychometric test only through the cognitive ability that it is supposed to assess.

\subsubsection{Example: link with the mixed-effect modeling}\label{mixed-effect}
A standard approach in the case $Y$ has a continuous state space is to model its observations through a mixed-effect model.
We can take as an example the system $\bX=(D,Y,V,G,U)$, taking the case where in the true law, the intensities of $Y$ and $D$ are given by Equations (\ref{eq:dynY}) and (\ref{eq:dynD}), and where the observations of $Y$ are given by Equation (\ref{eq:Z}).
If death could be treated as an ignorable drop-out, that is if CAR(DYN) holds, as it does in the ``chicken'' example, a linear mixed effect model for the $Z_j$ (in which $Y_0$ can itself be modeled using fixed and random effects) could be used for estimating the ``causal parameter'' $\beta_2^*$.

 In human epidemiology, it is not likely to hold. It is more likely that either $Y$ itself influences Death, or a random effect $U$ influences both $Y$ and Death. In these more realistic cases, treating Death as drop-out leads to informative drop-out (CAR(DYN) does not hold).

\section{Illustration: effect of blood pressure on death and cognitive ability}\label{sec:Illustration}
There is a great interest in modeling the evolution of cognitive ability with ageing, and it is more and more recognized that death should be taken into account in such modeling.  Recent works have devised a joint model between a quantitative marker \citep{dantan2011joint} or a latent process \citep{Rouanet2016joint,proust2016joint} representing cognitive ability, and a multistate model featuring both death and dementia. We give here a representation of this kind of model in our framework. One difference with the above cited works comes from the clear distinction that we make between system and observation. In particular we consider, as in \cite{ganiayre2008latent}, that psychometric tests as well as diagnosis of dementia are but observations of the cognitive ability.

 We examine the enriched model $\bX^{\tilde M}$ for the effect of blood pressure on death and cognitive ability.
 We take the case where in the true law the processes $D$ and $Y$ have the intensities:
\begin{eqnarray}
 \lambda_{Yt}&=&\beta^*_0(t)+\beta^*_1 G+\beta^*_2 V_t +\beta^*_3U \label{eq:dynY}\\
 \lambda_{Dt}&=&1_{\{D_{t-}=0\}}\alpha^*_{0D}(t)e^{\gamma^*_1 G+\gamma^*_2 V_t+\gamma^*_3Y_t+\gamma^*_3U}\label{eq:dynD},
\end{eqnarray}

We assume that $V$ is observed exactly in continuous time. We assume that $D$ is observed in continuous time with possible right-censoring, so that we observe $(\tilde D,\delta)$, the possibly censored death time and the censoring indicator. We find the observations $Y_{tj}$ by integrating the differential equation up to $t_{j}$; we take the case where the martingale in the Doob-meyer decompsotion of $Y$ ia a Brownian:
$$Y_{t_j}=\int_0^{t_j}\lambda_{Yu} \dd u + B_{t_{j}}.$$
 Adding a measurement noise, we obtain the equation for the observed $Z_j$s under a Model-a:
\begin{equation}\label{eq:Z} Z_j=Y_{t_j}+\varepsilon_j=Y_0+ B_0^*(t_j) + \beta^*_1Gt_j + \beta_2^* \int_0^{t_j} V_u~~\dd u + \beta_3^*Ut_j + B_{t_j} + \varepsilon_j,\end{equation}
Where $B_0^*(t_j)=\int_0^{t_j} \beta^*_0(t)\dd t$ and where $\beta^*_2$ is  a ``causal parameter'' with the assumption that $\bX^{\tilde M}$ is (nearly) perfect; in this case $\beta^*_2$ characterizes the conditional effect.

  More realistically, $Y$ can be treated as a latent process. It can be indirectly observed through one or several psychometric tests and also diagnosis of dementia. A rather general threshold model was proposed by \citet{ganiayre2008latent} for representing the link between psychometric tests and dementia, and the latent cognitive ability; this was a type-b observation model (see Section \ref{sec:latent process}). We present a simplified version of it, still omitting the subscript $i$ indexing the subject.
We denote $Z_{j}^{1}$ the
random variable representing the observation of the psychometric test (which can be the Mini Mental State Examination: MMSE) on the occasion of
the $j^{th}$ visit at time $t_{j}$. We consider a test for which $Q$ ordered values are possible ($q \in
[0,Q-1] $). Observation of
$Z_{t_j}^1=q$ provides the information that $Y_{t_j}+\varepsilon^1_{j}$ lies between
two thresholds, that is:
\begin{equation} \label{eq:thresholdMMS} Z_{t_j}^1 = q ~~\mbox{ if and only if } ~~ c^{1}_{q} \le  Y_{t_j}+\varepsilon^1_{j} < c^{1}_{q+1},\end{equation}
 with $c^1_0=-\infty$ and $c^1_{Q}=+\infty$.  The cut-off points $c^{1}_{q}$ are not known and so, are parameters to be estimated. 
The  $\varepsilon^1_{j}$s may be assumed to have a normal or a logistic distribution.

Binary data, such as diagnosis of dementia, are simply a special case of ordinal data for which we
only need one cut-off point, $c^2$ for instance:
\begin{equation} \label{eq:thresholdDementia} Z_{t_j}^2  =1_{\{Y_{t_j}+\varepsilon^2_{j} \ge c^2\}}.\end{equation}
Table \ref{Table:Obssystemcog} summarizes the observation mechanism of the system.

 The parameters can be estimated assuming a joint model with the same structure as Equations (\ref{eq:dynY}) and (\ref{eq:dynD}), in which case the model is well specified.
 We may assume that the RIPs for both $Y$ and $D$ are CAR(DYN), which allows us to avoid modeling them. For writing the likelihood (ignoring the mlid) we first write the likelihood conditional on the random effect $U$, for the observation $Z$ of $Y$, $\LL_{Z|U}$, and for the observation $\check D=(\tilde D,\delta)$ of $D$ given $U$, $\LL_{\check D|U}$; then we integrate out the random effects to compute $\Ee(\LL_{Z|U}\LL_{\check D|U})$.
Subject to identifiability, the maximum likelihood estimators are consistent. The main difficulties are numerical. One of the difficulties comes from the threshold model, and another from the integration over random effect;  \citet{proust-lima_LCMM2015} proposed continuous approximations of the step function $g$ for attenuating the former and latent class models for the latter (latent classes are in fact defined by random effects which can take a finite number of values).

\begin{table}
\caption{The observation of the system $\bX$ for cognitive ability. The most complex observation is for $Y$: a Model-b is used for linking $Y$ to observation of both MMSE and diagnosis of dementia at times $(t_1,\ldots,t_m)$.}\label{Table:Obssystemcog}
\begin{center}
\begin{tabular}{|l|lll|l|}
\hline
System  & Noise &  $g$ & RIP & Observed \\
\hline
Cognitive ability: $Y$ & $\varepsilon^1$ &  eq (\ref{eq:thresholdMMS}) & $R^1_{Yt}=1_{t\in (t_1,\ldots,t_m)}$ & MMSE\\
 & $\varepsilon^2$ &  eq (\ref{eq:thresholdDementia}) & $R^2_{Yt}=1_{t\in (t_1,\ldots,t_m)}$ & Dementia \\
Blood pressure: $V$ &   & I & $R_V=1$&  $V$\\
Attribute: $G$&   & I & $R_{G}=1$&  $G$\\
Attribute: $U$&  & I & $R_{U}=0$& unobserved\\
Death: $D$ &   & I & $R_D=1_{C<t}$& $(R_D,R_DD)$\\

\hline

\end{tabular}
\end{center}
\end{table}

\section{Conclusion}\label{sec:conclusion}
We have proposed a general approach based on the stochastic system approach to causality to study the effect of a factor on a process of interest taking binary or quantitative values, when the risk of death in non-negligible. We have argued that in that case one must first look at the effect of the factor on death, then on the process of interest. We have studied different observation schemes making a typology of cases where the mechanism leading to missing data was ignorable. We have examined the cases where ignoring death could lead to unbiased estimates of the effect on the process of interest. Finally we have illustrated this approach in analysing the structure of the system and its observation in a study of the effect of blood pressure on cognitive performance in the elderly.
All that has been said is valid when $V$ is a process that is completely observed or an observed fixed variable, which would then be considered as an attribute. Often, physiological markers, like blood pressure, are treated as fixed variables; this is of course a crude  approximation because blood pressure can vary in time.

We think that this approach can give a framework for analysing the evolution of physiological and pathological processes in epidemiology, and that this approach can also be applied to other fields.\vspace{5mm}

{\bf Acknowledgement:} I thank H\'el\`ene Jacqmin-Gadda for interesting discussions and comments, and the MELODEM initiative for raising interesting questions.


\label{lastpage}

\end{document}